\documentclass[useAMS,usenatbib]{mn2e}

\usepackage{psfig, epsf, epsfig}

\title[Group-group merging]{The formation of the NGC 1052 galaxy trail
via group-group merging}
\author[K. Bekki,   D. Vaz, D. A. Forbes, W. J. Couch]
{Kenji Bekki${}^1$\thanks{E-mail:
kenji.bekki@uwa.edu.au},
Daniel Vaz${}^2$, Duncan  A. Forbes${}^2$, 
and Warrick, J.  Couch${}^2$ \\
${}^1$ICRAR M468
The University of Western Australia
35 Stirling Hwy, Crawley
Western Australia 6009, Australia \\
${}^2$ Center for Astrophysics and 
Supercomputing, Swinburne University, John Street, Hawthorn VIC 3122, Australia
}

\begin{document}

\date{Accepted, Received 2005 February 20; in original form }

\pagerange{\pageref{firstpage}--\pageref{lastpage}} \pubyear{2005}

\maketitle

\label{firstpage}

\begin{abstract}
We propose a new scenario for the formation of the galaxy trail---including 
ultra-diffuse galaxies (UDGs)---discovered in the NGC 1052 group. In this 
scenario, a smaller group consisting of more than 10 gas-rich dwarf galaxies 
merges with the NGC 1052 group, ultimately forming a narrow stream 
of galaxies 
originating from the infalling group. We numerically investigate the 
evolution 
of gas-rich dwarf galaxies during this group merger, placing a strong 
emphasis 
on the formation of massive star clusters (SCs)
and dark matter fractions ($f_{\rm dm}$). 
The principal results of our new models are as follows:
The rapidly varying, time-dependent gravitational potential of the group 
merger can transform extremely gas-rich dwarf disk galaxies into UDGs within 
a timescale of 4~Gyr. 
Strong tidal compression of the gas disks induced by the group--group 
merger can dramatically enhance the formation of massive young SCs in these 
gas-rich dwarfs. 
More compact dwarf galaxies are less strongly affected by the time-varying 
gravitational potential, eventually evolving into nucleated dwarf galaxies 
dominated by dark matter. 
$f_{\rm dm}$ of the transformed galaxies 
depend on the initial structures of their dark matter halos. 
Depending on the initial properties of the dwarf galaxies, this scenario 
can produce UDGs with low dark matter fractions and 
populations of overluminous SCs. 
We predict that the galaxy trail should contain not only  compact, 
low-mass galaxies dominated by dark matter halos, but also ultra-faint dwarf 
galaxies. We further predict that dark matter--deficient, UDGs with massive SCs should 
be present in other environments that have experienced group mergers.

\end{abstract}


\section{Introduction}

Since ultra-diffuse galaxies (UDGs) were discovered across diverse galaxy 
environments (e.g., van Dokkum et al. 2015; Koda et al. 2015; Rom\'{a}n et al. 2017; 
Janssens et al. 2017; van der Burg et al. 2017; Forbes et al. 2020a), the physical origins of their 
characteristic properties have been investigated both observationally and 
theoretically (see Gannon et al. 2026 for a recent review). Although recent 
numerical and theoretical studies have demonstrated that various physical 
processes may drive UDG formation---including early gas stripping via ram 
pressure (e.g., Yozin \& Bekki 2015; Tremmel et al. 2020), high halo spin 
parameters (e.g., Amorisco \& Loeb 2016; Rong et al. 2017), tidal interactions 
and stripping (e.g., Ogiya 2018; Carleton et al. 2019; Sales et al. 2019;
Jackson et al. 2021; Moneno et al. 2022), 
and galaxy mergers (e.g., Wright et al. 2021; Tapia et al. 2026)---it remains 
to be fully understood which mechanism dominates. Furthermore, the diverse 
physical properties among UDGs, such as variations in globular cluster 
(GC) richness and gas mass fractions,
suggest that they form through multiple 
evolutionary pathways (e.g., Forbes et al. 2020b; Zaritsky et al. 2025).

\begin{table*}
\centering
\begin{minipage}{170mm}
\caption{
Initial model parameters and final properties for the selected dwarf galaxy
models (M1--M7): model ID (1st Column), adopted initial dark matter halo profile (2nd),
dark matter halo mass ($M_{\rm dm}$; 3rd), stellar mass
($M_{\rm s}$; 4th),  gas mass ($M_{\rm g}$; 5th),
stellar disk size ($R_{\rm s}$; 6th), final stellar half-mass radius
($R_{\rm h}$; 7th), final mean $g$-band surface brightness within $R_{\rm h}$
($\langle \mu_{\rm g} \rangle_{\rm h}$; 8th), 
final dark matter mass fraction within $R_{\rm h}$
($f_{\rm dm}$; 9th), and comments on the final properties of the simulated
galaxy (10th).
}
\begin{tabular}{cccccccccc}
ID
& Halo  
& $M_{\rm dm}$ (${\rm M}_{\odot}$) 
& $M_{\rm s}$ (${\rm M}_{\odot}$)
& $M_{\rm g}$ (${\rm M}_{\odot}$)
& $R_{\rm s}$ (kpc)
& $R_{\rm h}$ (kpc)
& $\langle \mu_{\rm g} \rangle_{\rm h}$ (mag arcsec$^{-2}$)
& $f_{\rm dm}$ 
& Comment  \\
\hline
M1 & cored & $6.7 \times 10^{10}$ &  $2.0 \times 10^8$ & $4.0 \times 10^8$ 
& 4.0 & 2.2 & 24.8 & 0.42 &  SC-rich UDG     \\
M2 & cusp & $1.7 \times 10^{10}$ &  $1.0 \times 10^8$ & $1.0 \times 10^8$ 
& 1.5 & 0.9 & 23.4 & 0.81 & Compact dwarf    \\
M3 & cusp  & $1.7 \times 10^{10}$ &  $1.0 \times 10^7$ & $4.0 \times 10^7$ 
& 2.0 & 1.0 & 26.3 & 0.84 & Nucleated   \\
M4 & cored & $5.0 \times 10^{10}$ &  $5.4 \times 10^8$ & $6.0 \times 10^7$ 
& 4.0 & 1.1 & 23.3 & 0.04 & UCD in a stream    \\
M5 & cusp & $6.7 \times 10^{10}$ &  $2.0 \times 10^8$ & $4.0 \times 10^8$ 
& 4.0 & 2.4 & 25.1 & 0.86 & DM-dominated   \\
M6 & cored & $6.7 \times 10^{10}$ &  $2.0 \times 10^8$ & $4.0 \times 10^8$ 
& 4.0 & 1.9 & 23.9 & 0.70 & SC-rich dwarf   \\
M7 & cusp & $5.0 \times 10^{10}$ &  $1.5 \times 10^8$ & $3.0 \times 10^8$ 
& 4.0 & 1.3 & 23.9 & 0.49 & Nucleated     \\
\end{tabular}
\end{minipage}
\end{table*}

van Dokkum et al. (2022) recently discovered an almost straight-line 
trail of $\sim 10$ faint galaxies, including the UDGs DF2, DF4, and DF9, 
in the NGC 1052 group. An intriguing observational result regarding these 
galaxies is that at least DF2, DF4, and  DF9 are deficient in dark matter
(e.g., van Dokkum et al. 2018a, Emsellem et al. 2019; Keim et al. 2026). Both 
the observed galaxy trail and the dark matter deficiency in DF2 and DF9 
have been considered to be explained self-consistently by the ``bullet-dwarf 
collision'' scenario (Silk 2019), in which two gas-rich, dark 
matter--embedded dwarfs collided head-on approximately 8~Gyr ago to form 
new, dark matter--free faint galaxies. 
Recent hydrodynamical simulations have demonstrated that a trail of 
new dark matter--free dwarf galaxies can indeed form in a head-on collision 
between extremely gas-rich dwarf galaxies (e.g., Shin et al. 2020; Lee et al. 2024).

Although the bullet-dwarf collision scenario appears promising, it does not 
clearly explain why the initial straight-line distribution of the newly formed 
dwarfs has been preserved around the NGC 1052 group over many gigayears. 
Furthermore, Vaz et al. (2026) recently discovered a dark matter--dominated 
low-mass dwarf galaxy (LEDA 4014647) within the NGC 1052 trail,
which is hard to be explained by the bullet-dwarf collision scenario. These issues suggest that an alternative 
scenario must be considered for the formation of the NGC 1052 trail.

We hypothesize that the trail formed from the ancient merger of the NGC 1052 
group with a smaller infalling group. In this group--group merging scenario 
(hereafter GGM), the member galaxies of the smaller group are tidally stripped 
during the merger, ultimately distributing themselves into a narrow stream 
that is observed as a ``trail'' around NGC 1052. It remains unclear, however, 
why some member galaxies become dark matter--deficient UDGs (e.g., DF2 and DF4) 
while others become compact, dark matter--dominated dwarfs 
within this GGM framework.

The purpose of this paper is thus to investigate whether the GGM scenario 
can explain both the origin of the NGC 1052 trail and the physical properties 
of its constituent galaxies, based on numerical simulations of dwarf galaxy 
evolution during group mergers. Specifically, we examine how gas-rich dwarf 
galaxies initially residing in the smaller infalling group are transformed 
into other galaxy types (e.g., UDGs), and whether the formation of the observed 
overluminous 
star clusters (SCs; van Dokkum et al. 2018b) is enhanced within these galaxies 
during the merger. 

\section{Models}

We adopt our original simulation code (Bekki 2015) to perform 
$N$-body/SPH simulations of dwarf galaxy evolution during group-group merging
(GGM), incorporating 
morphological and dynamical evolution, star formation histories, star cluster 
(SC) formation, and chemical enrichment. In doing so, we follow a two-step 
procedure: First, we investigate the time evolution of the spatial distribution 
of dwarf galaxies initially residing in the smaller group during the GGM event. In this 
first step, we aim to identify a ``good'' GGM model in which the final distribution 
of galaxies forms a narrow stream as observed, by exploring models with various 
mass ratios and orbits for the merging groups. To clearly trace the spatial 
distribution of the dwarfs during GGM, we assume in this step that only the 
smaller group hosts dwarf galaxies; each member galaxy is represented by a 
collisionless point-mass particle (a ``galaxy particle'').

Once a good GGM model is identified in the first step, we replace one of the 
galaxy particles with a live, gas-rich dwarf galaxy model composed of $N$-body/SPH 
particles and rerun the simulation. In this second step, we perform full 
$N$-body/hydrodynamical simulations of the dwarf galaxy within the good GGM 
model, repeating this process for multiple individual galaxy particles. By 
combining these two steps, we can simultaneously investigate both the large-scale 
final spatial distribution of galaxies after GGM and the internal evolution of 
individual dwarf galaxies. The mass and spatial resolutions in the first 
collisionless step are $5 \times 10^6 \, {\rm M}_{\odot}$ and 3~kpc, respectively, 
which are sufficient to resolve the large-scale spatial distribution of galaxies 
around NGC 1052. In the second step, the mass and spatial resolutions for
newly formed stars are $6 \times 10^2 \, {\rm M}_{\odot}$ and 2.3~pc, respectively, 
allowing us to resolve the formation of massive SCs. Multiple gravitational 
softening lengths for different components (Bekki 2015) are adopted to suppress 
artificial dynamical heating of the dwarf disks by massive dark matter particles.

The dark matter halos of both merging groups are represented by the ``NFW model'' 
with central cusps (Navarro, Frenk \& White 1996), as predicted by cold dark 
matter (CDM) simulations. The total dark matter mass ($M_{\rm gr,1}$), virial 
radius ($R_{\rm vir,1}$), and concentration parameter ($c$) of the larger group 
(corresponding to the NGC 1052 group) are set to $5 \times 10^{12} \, {\rm M}_{\odot}$, 
300~kpc, and 7, respectively, all of which are consistent with observations of 
the NGC 1052 group (e.g., Forbes et al. 2019). We explore various GGM models 
with different masses ($M_{\rm gr,2}$) and virial radii ($R_{\rm vir,2}$) for 
the smaller group. We assume that: (i) the smaller group initially contains 
15 dwarf galaxies,
(ii) their spatial distribution follows the dark matter 
density profile of the group, and (iii) their distribution is truncated at 
$0.5 R_{\rm vir,2}$.

The initial 3D positions $(x, y, z)$ and velocities $(v_{\rm x}, v_{\rm y}, v_{\rm z})$ 
of the smaller group relative to the larger group (which is initially at rest) 
are set to $(1.3 R_{\rm vir,1}, b R_{\rm vir,1}, 0)$ and $(-f_{\rm v} v_{\rm c}, 0, 0)$, 
respectively. Here, $b$ is the impact parameter of the GGM, $v_{\rm c}$ is the 
circular velocity at $R_{\rm vir,1}$ of the larger group, and $f_{\rm v}$ is a 
dimensionless parameter governing the orbit of the smaller group. We have 
explored a grid of models with varying group mass ratios ($m_2 = M_{\rm gr,2}/M_{\rm gr,1}$), 
size ratios ($r_2 = R_{\rm vir,2}/R_{\rm vir,1}$), $b$, and $f_{\rm v}$, identifying 
several good models that clearly produce a galaxy trail. For example, as shown 
in Fig.~1, the model with $m_2 = 0.1$, $r_2 = 0.3$, $b = 0.1$, and $f_{\rm v} = 1.1$ 
forms a narrow stream of galaxies within $\approx 5$~Gyr of the group merger, 
appearing as a ``trail'' around NGC 1052. 
It is found that only models with lower $m_2$ ($<0.2$) can show such a trail 
in the present study.

The total masses of the dark matter halo, stellar disk, and gas disk of a dwarf 
disk galaxy are denoted as $M_{\rm dm}$, $M_{\rm s}$, and $M_{\rm g}$, respectively. 
Because Bekki et al. (2003) demonstrated that the dynamical evolution of dwarfs 
in galaxy groups and clusters depends strongly on whether their dark matter halos 
follow the ``cuspy'' profile or the ``cored'' profile proposed by Salucci \& 
Burkert (2000), we investigate both dark matter halo models. The radial ($R$) 
and vertical ($Z$) density profiles of the stellar disk (with truncation radius 
$R_{\rm s}$) are proportional to $\exp(-R/R_{0})$ with scale length $R_{0} = 0.4 R_{\rm s}$ 
and ${\rm sech}^2(Z/Z_{0})$ with scale height $Z_{0} = 0.04 R_{\rm s}$, respectively. 
We examine models in which the gas disk scale length ($R_{\rm 0,g}$) is set to 
either $R_{0}$ or $2 R_{0}$ in order to evaluate how our results depend on 
$R_{\rm 0,g}$. 

An SPH gas particle is converted into a new stellar particle at each time step 
when the local gas density exceeds $100 \, {\rm atoms \, cm}^{-3}$. To identify 
young massive SCs at a given time step, we compute the total mass of newly formed 
stars within 20pc around each new stellar 
particle. If this enclosed mass exceeds $10^5 \, {\rm M}_{\odot}$, we designate 
the central stellar particle as a massive SC and investigate  the total  mass 
($M_{\rm sc}$)
and the mean mass density within 10pc ($\rho_{\rm sc}$).
Chemical enrichment and feedback from supernovae 
and asymptotic giant branch (AGB) stars are modeled 
and the initial iron abundance of 
the dwarf galaxy gas disk is set to a fiducial value of ${\rm [Fe/H]} = -1.6$, 
though different initial ${\rm [Fe/H]}$ values are explored across the model suite.

Although we have simulated a wide suite of galaxy models spanning different values 
of $M_{\rm dm}$, $M_{\rm s}$, $M_{\rm g}$, and orbital parameters, we primarily 
focus on the results of our fiducial model, M1, which clearly demonstrates the 
formation of a dark matter--deficient, SC-rich UDG during GGM. Table~1 summarizes 
the model parameters and final galaxy properties for the seven selected models 
discussed in this paper; results from the broader model suite will be presented 
in a forthcoming paper. To determine whether a simulated galaxy can be classified 
as a UDG, we convert the stellar surface mass density ($\Sigma$) into a $g$-band 
surface brightness ($\mu_{g}$) assuming a constant $g$-band mass-to-light ratio 
($\Upsilon_{g}$) of 1.8 for all stars. Finally, the stellar half-mass radius ($R_{\rm h}$), 
the dark matter mass fraction ($f_{\rm dm}$) within $R_{\rm h}$, and the mean 
$g$-band surface brightness within 
$R_{\rm h}$ ($\langle \mu_{\rm g} \rangle_{\rm h}$) are measured 
at the final time step.

\begin{figure} 
\psfig{file=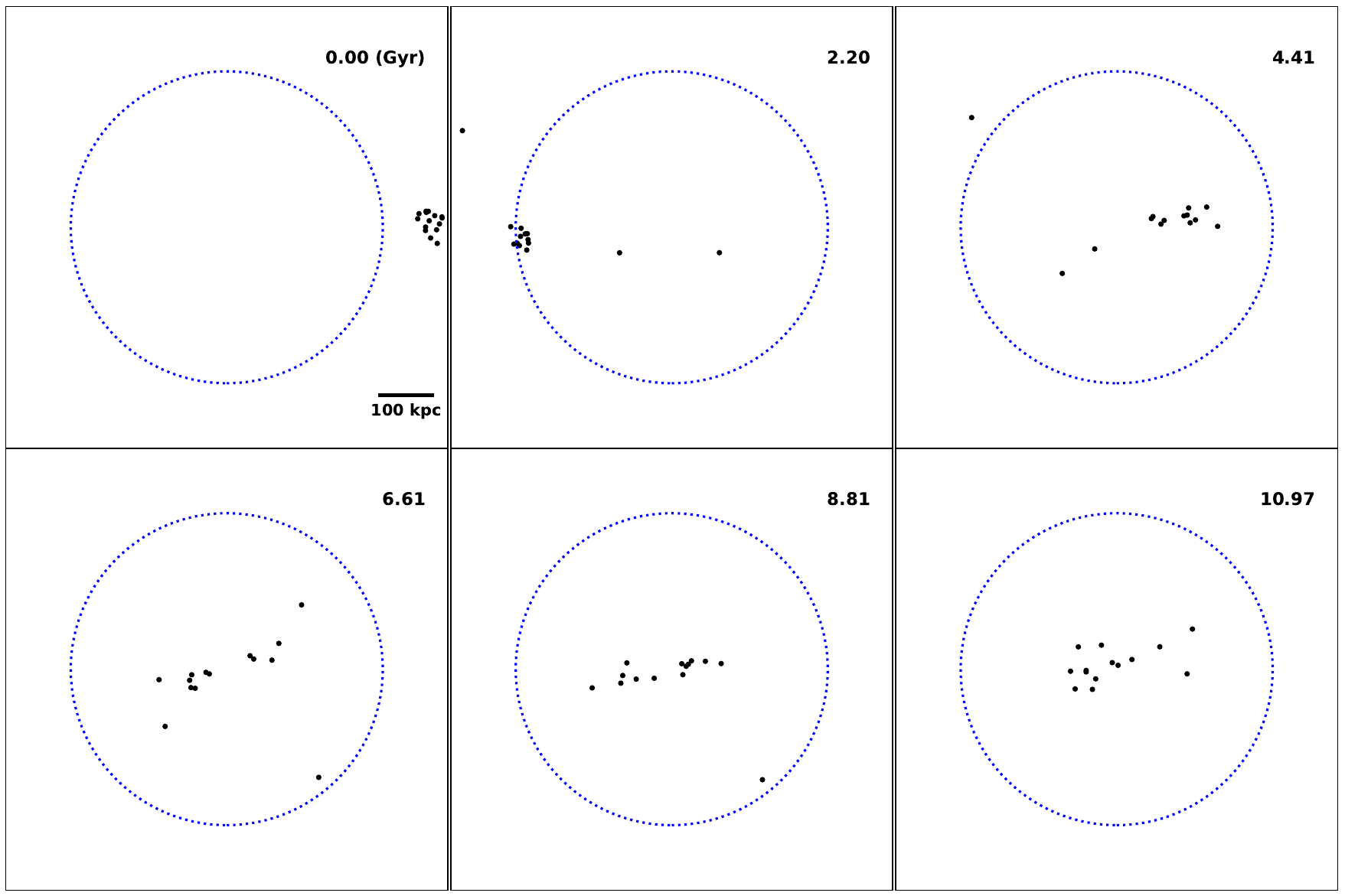,width=8.0cm}
\caption{
Spatial distributions of the 15 dwarf galaxies initially residing in the 
smaller infalling group, projected onto the $x$--$z$ plane (upper two panels) 
and the line-of-sight velocities ($V_{\rm y}$) as a function of the $x$-position 
(lower two panels) at $T = 0$~Gyr and $5.53$~Gyr. Each galaxy is represented 
by a small  black or green star, and
the central galaxy NGC 1052 is indicated by a red star. 
We investigate galaxy transformation processes along the orbit of the green particle.
The virial radius ($R_{\rm vir} = 300$~kpc) of the NGC 1052 group dark matter halo 
is shown by the blue dotted circle. For illustrative purposes, a light-blue band 
of width $60$~kpc
and two dotted blue lines separated by a velocity difference of 
100 km s$^{-1}$ are included in the upper and lower right panels, respectively. 
These can be compared with recent observations by Keim et al. (2025). 
This GGM  model, with $M_{\rm gr,1} = 5 \times 10^{12} \, {\rm M}_{\odot}$, 
and $m_2 = 0.1$,  serves as a representative example of GGM models 
that successfully reproduce a galaxy trail around the NGC 1052 group.
}
\label{Figure. 1} 
\end{figure}

\begin{figure} 
\psfig{file=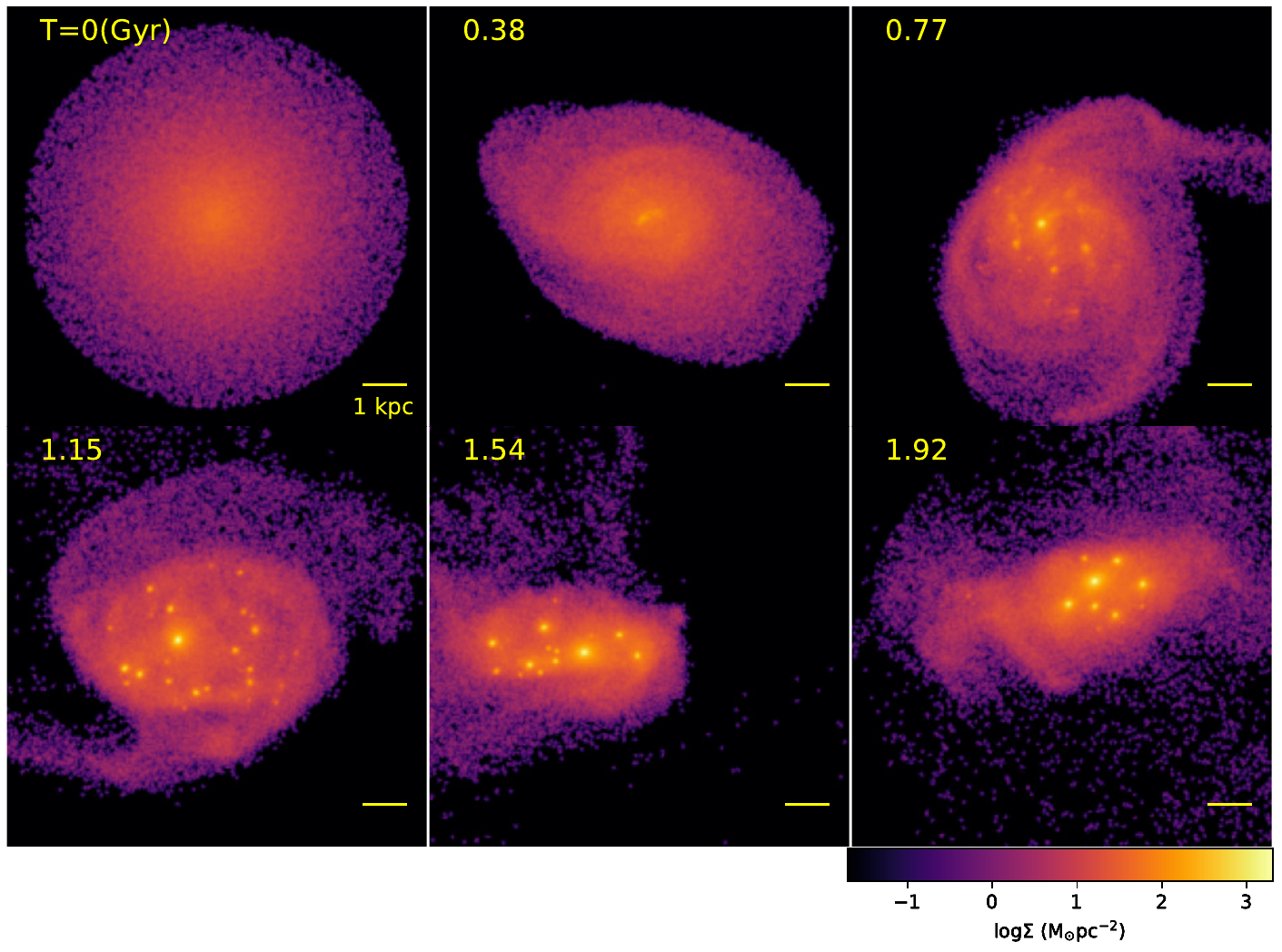,width=9.0cm}
\caption{
Time evolution of the stellar surface mass density ($\Sigma$) projected onto
the $x$--$y$ plane for the gas-rich dwarf galaxy model M1. A scale bar of 1~kpc
and the simulation time ($T$) are shown in the lower-right and upper-left
corners of each panel, respectively. The $15~{\rm kpc} \times 15~{\rm kpc}$
region is divided into $200 \times 200$ grid cells, with $\Sigma$ evaluated
at each grid point. Massive young SCs are clearly formed via tidal compression
of the gas-rich dwarf during GGM.
}
\label{Fig. 2} 
\end{figure} 

\begin{figure} 
\psfig{file=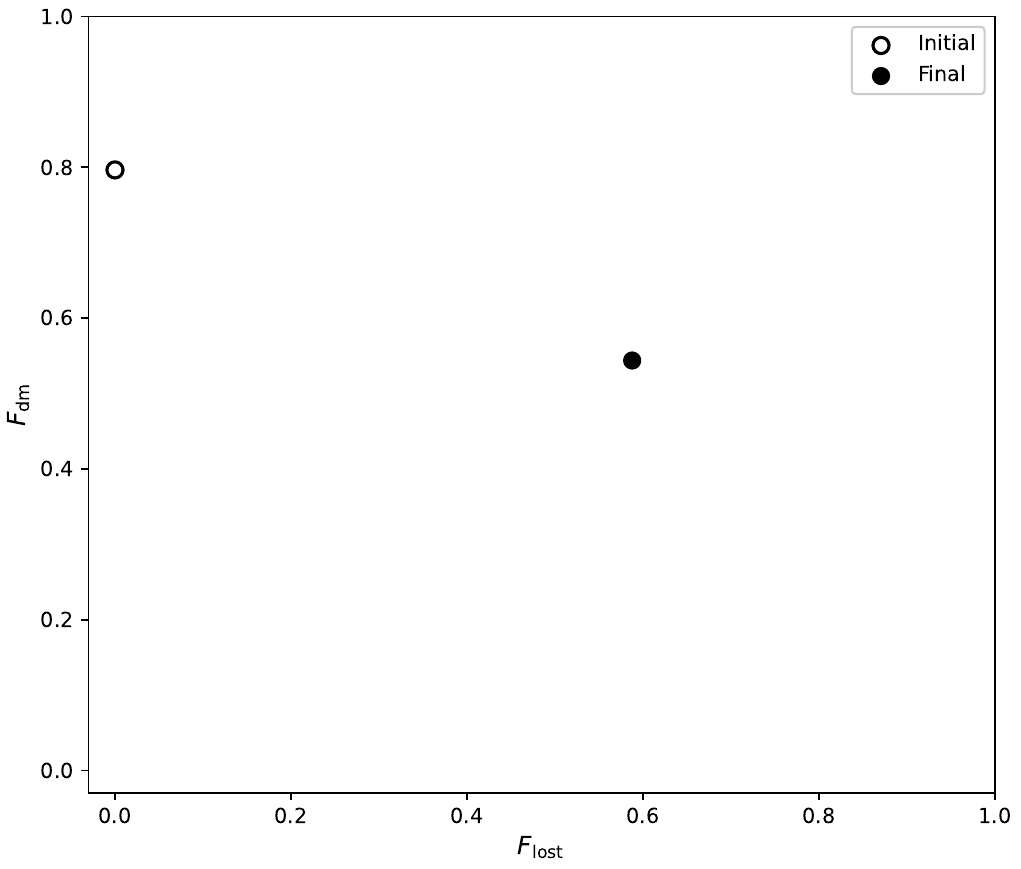,width=8.0cm}
\caption{
Final 2D maps of the stellar surface mass density ($\Sigma$) for the six
additional galaxy models (M2--M7). 
The formation of massive SCs is clearly
visible in M3-M7.
M4 is a UDG-UCD transition object.
}
\label{Fig. 3} 
\end{figure}

\section{Results}

Fig.~2 illustrates how an extremely gas-rich dwarf disk galaxy is transformed 
into a UDG hosting multiple massive SCs during GGM in the fiducial model M1. 
The initial disk morphology quickly disappears due to strong tidal heating 
by GGM ($T = 1.2$~Gyr), and intense starbursts are triggered during GGM
 with a peak star 
formation rate (SFR) of $0.2 \, {\rm M}_{\odot} \, {\rm yr}^{-1}$ at $T = 0.9$~Gyr 
and $3.3$~Gyr. Multiple high-density stellar clumps---the progenitors of massive 
SCs---form as a result of strong tidal compression of the gas disk induced by 
the time-dependent gravitational potential of the merger. Consequently, the 
dwarf galaxy develops 28 SCs 
with $M_{\rm sc} \ge  10^5 \, {\rm M}_{\odot}$, accounting for 
mass fractions of 0.03 
among all newly formed stars.

The dwarf galaxy loses most of its initial dark matter due to strong tidal
stripping by the merging groups, causing the final 
dark matter fraction ($f_{\rm dm}$) within $R_{\rm h}$ to drop to $0.42$. 
The total masses of dark and baryonic matter  within 
$R_{\rm h}$  at $T = 3.85$~Gyr are $1.7 \times 10^8 \, {\rm M}_{\odot}$, 
and $2.3 \times 10^8  \, {\rm M}_{\odot}$, respectively. 
The line-of-sight velocity dispersions ($\sigma$) of stars at
$R_{\rm h}$, projected onto the $x$--$y$, $x$--$z$, and $y$--$z$ 
planes, are 18, 13, and 24 km s$^{-1}$, respectively. This implies (i)
a factor of $\approx 4$ uncertainty in dynamical mass estimates derived from 
$\sigma$ and $R_{\rm h}$ alone and (ii)
the dark matter mass can be underestimated for certain viewing angles, as proposed by Montes et al. (2021) for DF2.
The mean $g$-band 
surface brightness within $R_{\rm h}$ is $\approx 25 \, {\rm mag \, arcsec}^{-2}$, 
indicating that this galaxy would be classified as a dark matter--deficient, 
GC-rich UDG once its SCs evolve into GCs.
The simulated tidal features (e.g., streams and tails) suggest that 
stellar streams, such as those observed near some  UDGs
(e.g., NGC 2708-Dw1; Jones et al. 2021), may be found within the 
NGC 1052 trail.

Fig.~3 demonstrates that GGM produces substantial morphological diversity 
depending on the initial properties of the infalling galaxies. For example, 
a low-mass, compact dwarf galaxy is less susceptible to tidal stripping by 
GGM and instead evolves into an actively star-forming galaxy that remains 
dominated by dark matter (M2). This suggests that LEDA 4014647 in the NGC 1052 
group was able to retain a significant fraction of its initial dark matter 
following its infall due to its compact initial configuration. It further 
implies that other compact dwarf galaxies in the group, such as RCP 16 and 
RCP 28, are also likely dominated by dark matter. Nucleated dwarfs/UDGs 
(like DF9)  can also 
form when newly formed SCs spiral into the central regions of their progenitor 
dwarf galaxies (M3 \& M7), though some of these systems subsequently transform into 
ultra-compact dwarf galaxies (UCDs) following the nearly complete tidal 
stripping of their host galaxies (M4). The dark matter fractions of these 
nucleated galaxies vary considerably depending on their orbits and dark matter
profiles.

More compact, gas-rich dwarf galaxies with higher initial dark matter densities 
transform into UDGs containing numerous SCs without losing a substantial fraction 
of their dark matter (M5, adopting the NFW halo model). This indicates that 
SC-rich (and hence GC-rich) UDGs produced via GGM can be either dark matter--rich 
or dark matter--deficient, depending on their initial dark matter density profiles. 
More massive and compact  models yield higher mass fractions of young massive SCs 
(M6), demonstrating that initial galaxy mass is a key factor governing 
SC formation during GGM. Furthermore, while SCs can form in models with lower 
initial mass densities, their host dwarf galaxies are completely disrupted 
within $5$~Gyr. 
As suggested by Beasley \& Trujillo (2016),
such stripped SCs from dwarfs
would be observationally identified as intragroup GCs or UCDs 
(e.g., Buzzo et al. 2023).

Fig.~4 shows that high-density SCs 
with $\rho_{\rm sc} \ge 10\,{\rm M}_{\odot}\,{\rm pc}^{-3}$ 
within $10$~pc, which may survive tidal disruption, are systematically more massive, 
with a mean mass of $7.0 \times 10^5\,{\rm M}_{\odot}$.
Simulated SCs with lower densities and lower masses are more likely to be 
disrupted during their dynamical evolution.
The formation of high-density 
SCs is enabled by the very high initial gas fractions ($\ge 0.5$) assumed for the 
dwarf galaxies in this study. This implies that the infall of the smaller group 
into the NGC 1052 group must have occurred at an early cosmic epoch when its 
member galaxies were still very gas-rich. If all GCs in the UDG NGC 1052-DF2 
formed via GGM, their observed ages (e.g., Fahrion et al. 2025) suggest that 
the merger took place approximately $9$~Gyr ago. Moreover, the wide observed 
range of metallicities ($-1.86 \le {\rm [M/H]} \le -0.73$) in DF2 (Fahrion et al. 2025) 
is naturally explained by the GGM scenario, wherein gas-rich dwarf galaxies 
undergo sustained chemical enrichment driven by merger-induced starbursts.

\begin{figure} 
\psfig{file=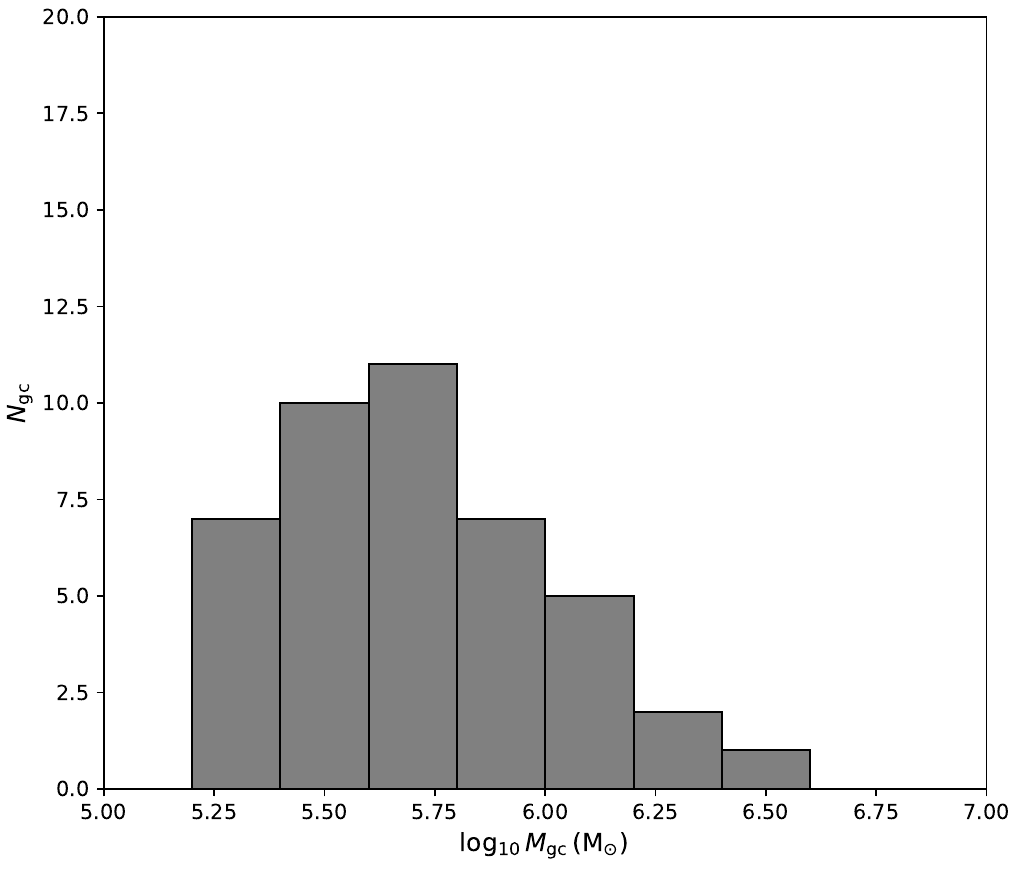,width=8.0cm}
\caption{
Normalized SC numbers ($N_{\rm sc}$) as a function of cluster mass 
($M_{\rm sc}$) for all SCs (blue) and high-density SCs (red) with 
$\rho_{\rm sc} \ge 20\,{\rm M}_{\odot}\,{\rm pc}^{-3}$ within $10$~pc, formed 
across models M1, M4, M5, M6, and M7. The vertical blue and red dotted lines 
indicate the mean masses for all SCs and high-density SCs, respectively.
}
\label{Fig. 4} 
\end{figure} 

\section{Discussion and conclusions}

We have first demonstrated that the GGM scenario can not only explain 
the galaxy trail in the NGC 1052 group but also reproduce the formation 
of  UDGs with massive SCs 
that are deficient in dark matter. Furthermore, the GGM 
scenario provides the following predictions regarding the properties of 
galaxies in the NGC 1052 group and UDGs in other environments.
First, the UDGs DF2, DF4, and DF9 should  not be  completely dark matter--free; 
instead, they retain small fractions of dark matter. 
Second, the NGC 1052 trail should contain both dark matter--deficient UDGs 
and compact, dark matter--dominated galaxies. This prediction is supported 
by the recent discovery of a compact dwarf galaxy, LEDA 4014647, which is 
dominated by dark matter (Vaz et al. 2026), although the dark matter content 
of other compact dwarfs  remains to be investigated. 
The predicted coexistence of 
dark matter--deficient and dark matter--dominated 
galaxies stands in striking contrast to other scenarios 
(e.g., Lee et al. 2024; Hammer \& Yang 2026)
in which all dwarf galaxies in the stream should be almost 
completely dark matter--free.

Third, diffuse stellar streams should be present along the trail due to 
the tidal stripping of stars from the member galaxies of the merging smaller 
group. 
Fourth, dark matter--deficient UDGs should also exist in other galaxy groups 
and clusters that have recently experienced the infall or merging of smaller 
groups containing gas-rich dwarf galaxies. Indeed, Buzzo et al. (2026) 
recently discovered dark matter--deficient UDGs (FCC224 and FCC240) in the Fornax 
cluster, which could have formed through the merging of the cluster with 
a smaller group.
Fifth, young UCDs can form from nucleated UDGs 
through the tidal disintegration of their host galaxies. Although UCDs 
have previously been proposed to form via ``galaxy threshing'' (Bekki et al. 2003), 
the formation mechanism of the stellar nuclei themselves was not specified 
in those models. The GGM scenario thus provides a unified mechanism for 
both nucleus formation in gas-rich dwarfs and the subsequent disintegration 
of their host galaxies.

The trail may contain a number of ultra-faint dwarfs (UFDs) with 
$M_{\rm s} \le 10^5\,{\rm M}_{\odot}$, primarily because small galaxy groups 
with $M_{\rm dm} \approx 10^{12}\,{\rm M}_{\odot}$, such as the Local Group and 
the Sculptor group, are observed to host UFDs (e.g., Kallivayalil et al. 2018; 
Martínez-Delgado et al. 2024). The central galaxy of the smaller group may 
have recently merged with NGC 1052, triggering a central starburst in NGC 1052. 
More accurate distance measurements to galaxies in the trail, such as those 
by Tang et al. (2026), will place tight constraints on the orbits of merging 
groups in the GGM scenario.
We plan to investigate whether mutual tidal interactions among group members, 
which are not included in the present study, accelerate the transformation of 
gas-rich dwarfs into GC-rich UDGs
while further reducing their dark matter content.

{}

\end{document}